\documentclass{aastex63}

\usepackage{longtable}
\usepackage{geometry} %Use [showframe=true] optionally
\usepackage{changepage}

\received{May 24, 2021}
\accepted{May 26, 2021}

\submitjournal{RNAAS}

\shorttitle{One is the loneliest number}
\shortauthors{Cifuentes et al.}

\begin{document}

\title{One is the loneliest number: multiplicity in cool dwarfs
}

\correspondingauthor{Carlos Cifuentes}
\email{ccifuentes@cab.inta-csic.es}

\author[0000-0003-1715-5087]{Carlos Cifuentes}
\author[0000-0002-7349-1387]{Jos\'e A. Caballero}
\affiliation{Centro de Astrobiolog\'ia (CSIC-INTA), Madrid, Spain}
\author{Sergio Agust\'i}
\affiliation{Lyce\'e Fran\c{c}ais, Madrid, Spain}

\begin{abstract}
\begin{adjustwidth}{1cm}{1.2cm}
Stars in multiple systems offer a unique opportunity to learn about stellar formation and evolution.
As they settle down into stable configurations, multiple systems occur in a variety of hierarchies and a wide range of separations between the components.
We examine {11} known and {11} newly discovered multiple systems including at least one M dwarf with the latest astrometric data from {\em Gaia} Early Data Release 3 (EDR3).
We find that the individual components of systems at very wide separations are often multiple systems themselves. 
\end{adjustwidth}
\end{abstract}


\section{Introduction} 

Systems of two or more stars bound by gravity, or multiple star systems, are a common by-product of the stellar formation process.
By analysing these systems we aim to understand the mechanisms behind the star formation and evolution, particularly of low-mass objects.
The competition of the system components for stable orbits eventually resolves into a hierarchical arrangement in hierarchical systems \citep{Duc13}.
When a system reaches dynamical stability, it organises hierarchically in nested orbits that can be treated individually as two-body problems \citep[see e.g.][]{Eva68,Tok14}. 
Evidence suggests that most (if not all) stars form in multiple systems and at least one component (usually the least massive) is ejected into a distant orbit during the pre-main sequence phase \citep{Duc13}.
Very close binary systems ($r<10$\,au) are not formed by fragmentation in situ, but produced from wider multiple systems \citep{Bat03}.
In many physical star systems, seemingly single distant components are resolved as multiple systems themselves, typically binaries, which implies that known double or triple systems in wide orbits would actually be hierarchical triples or quadruples.
This fact is in consonance with the suggestions by \citet{Bas06} and \citet{Cab07}, who proposed a major prevalence of wide triples over wide binaries.
These findings challenge the fundamentals of star formation because the separations of wide systems cannot be the direct product of a collapsing cloud core. 

Using N-body simulations, \citet{Rei12} found that extreme hierarchical architectures can be reached on timescales of millions of years with the appropriate exchange of energy and momentum.
In their work, one component in a triple system is dynamically scattered into a very wide orbit at the expense of shrinking the orbit of the remaining binary. 
Low-mass objects are preferentially ejected in three-body dynamics.
 \citet{Rei12} noted that these ejected companions can also be binaries.

In this work we examine the characteristics of known and new multiple systems containing at least one M dwarf by using the astrometry from {\em Gaia} EDR3.
In particular, we compare distances, proper motions, and radial velocities to determine if the systems are physically bound.

\section{Sample} 

We study {53} stars in {22} multiple systems, {11} of which are known systems tabulated in the Washington Double Star catalog \citep[][e.g. \citealt{Cab07,Cab12,Dhi10,Jan14}]{Mas01}, {4} are known systems with new candidate members, and {11} are newly discovered systems.
All of them contain at least one M dwarf in different hierarchical configurations, including 12 double, 5 triple and 4 quadruple systems, plus one quintuple system.
They are located between {14.1}\,pc and {276.1}\,pc.
The spectral classifications of the primary components range from F7\,V to M5.5\,V, and of their physical companions from F9\,V to L1.
We estimated spectral types for {22} stars from luminosities and absolute magnitudes, as described below. 

For each star, we acquired publicly available all-sky survey data and images with the Aladin interactive sky atlas \citep{Bon00}, and manipulated them with the Tool for OPerations on Catalogues And Tables \citep[TOPCAT;][]{Tay05}.
First, we obtained astrometry (equatorial coordinates in the J2016.0 epoch, parallaxes, and proper motions) and photometry ($G_{BP}$, $G$, and $G_{RP}$) from {\em Gaia} EDR3 \citep{Gaia21bro}.
%{\em Gaia} information is complete for all stars except {two}, for which we retrieved proper motions and parallactic distances from the literature \citep{Lep05,Dit14}.
Next, we compiled additional photometry in the $JHK_s$ passbands of the Two Micron All-Sky Survey \citep[2MASS;][]{Skr06} and in the $W1W2W3W4$ passbands of the Wide-field Infrared Survey Explorer \citep[AllWISE;][]{Cut14}.
%2MASS and AllWISE are complete for all except {four} and {ten} cases, respectively.
Using the photometry from up to 10 passbands and parallactic distances we derived luminosities and effective temperatures for the single stars as in \citet{Cif20}.
Finally, we used the mass-luminosity and mass-absolute magnitude relations by \cite{Pec13} and \cite{Cif20} to derive masses and spectral types.

\section{Multiplicity} 

In order to test the physical binding of each system member candidate, we quantified the similarity of distances and proper motions.
For each pair of objects we computed the proper motion ratio, $\Delta \mu / \mu$, and the proper motion position angle difference, $\Delta PA$, as in \cite{Mon18}, as well as the distance ratio, $\Delta d \equiv |d_A-d_B|/d_A$. 
We considered a pair as physical if all the following conditions are met:
$\Delta PA <$ 15\,deg,
$\mu$ ratio $<$ 0.15, and
$\Delta d <$ 0.10.
Of the {22} investigated systems, {7} do not comply with the criteria above.
Of them, {3} are close pairs with {\em Gaia} proper motions perturbed by their relative orbital motion.
% 1RXS J074948.5-031712 (32.82); WDS 20198+2257 (56.5); WDS 22259-7501 (489.2).
The remaining {4} pairs are wide: 
{one} does not have parallax measured by {\em Gaia} for the secondary (WDS~06104+2234), 
{one} exhibits a high Renormalized Unit Weighted Error (RUWE) value in {\em Gaia} (i.e. larger than 1.4, which is indicative of a problematic astrometric solution; KO~4), 
and only for the other {two} (KO~6 and SLW~1299) we disprove their physical connection based on the astrometric analysis.

In Table~\ref{table} we list the names of the {53} stars, spectral types, 
angular separations ($\rho$), 
position angles ($\theta$), 
projected physical separations ($s$),
and remarks.
The full version of this table with all astro-photometric parameters and derived masses, binding energies, orbital periods, and reference codes can be downloaded in csv format from our GitHub repository\footnote{\url{https://github.com/ccifuentesr/cif21-multiplicity} (10.5281/zenodo.4757508)}.
Additionally, one-page pdf charts with a detailed description of each system are also available in the repository.
The least-bound system in our sample is FMR~83 with $U^*_g \sim -7.1\,10^{33}$\,J \citep{RC12}, while the pair with largest projected separation (54\,800\,au) and orbital period (10.9\,10$^6$\,yr) is WDS~07400-0336 \citep{Mon18}.

\section{Results} 

We concluded the following:
($i$) In {6} multiple systems the astrometric data from {\em Gaia} suggest additional multiplicity in at least one member. % KO 1, KO 2, KO 4, 1RXS J073138.4+455718, HD 61606 A, 1RXS J074948.5-031712.
($ii$) With current data, the pairs LSPM~J0651+1845/LSPM~J0651+1843 and HD~77825/1RXSJ090406.8-155512 are the only wide binaries
%(112 and 220\,arcsec, respectively)
with no evidence of close binarity of the individual components in the {\em Gaia} solution.
($iii$) The young candidate member in $\beta$ Pictoris, PYC~J07311+4556, is a wide member in a quadruple, perhaps quintuple, physical system.
This current configuration is expected if the system is actually young \citep[$18.5^{+2.0}_{+2.4}$\,Myr,][]{MR20}
and still undergoes a process of dynamical stabilization.

\newpage

\begin{table}
\begin{adjustwidth}{-2.4cm}{}
\caption{Relative astrometry of our multiple systems. \label{table}}
\scriptsize
\begin{tabular}{llccccl}
\hline\hline
	{Star name} 	&	 {Comp.} 	&	 {Sp. type}  	&	 {$\rho$}  	&	 {$\theta$} 	&	 {$s$} 	&	 {Remarks} \\
  	{}  			&	 {}  		&	 {} 			&	{[arcsec]}  &	 {[deg]} 		&	 {[au]} 	&	 {} \\
\hline
LEHPM 494	&	A	&	m5.5\,V	&		&		&		&	Confirmed known binary (KO 1)	\\	
2MASS J00210589-4244433	&	B	&	L0.6:\,V	&	77.78	&	317.0	&	2083.4	&	B might be double	\\	\noalign{\smallskip}
NLTT 6496	&	A	&	M4.5\,V	&		&		&		&	Confirmed known binary (KO 4)	\\	
NLTT 6491	&	B	&	m4.5\,V	&	299.13	&	190.6	&	9391.9	&	B might be double	\\	\noalign{\smallskip}
LP 655-23	&	A	&	M4.0\,V	&		&		&		&	Confirmed known binary (KO 2)	\\	
DENIS J043051.5-084900	&	B	&	M8\,V	&	19.81	&	339.8	&	597.3	&	A and B might be double	\\	\noalign{\smallskip}
2MASS J06101775+2234199	&	A	&	M4.0\,V+	&		&		&		&	New triple. JNN 269 is visual 	\\	
LP 362-121	&	BaBb	&	M6\,V+m7\,V	&	65.16	&	89.2	&	1866.9	&	\\	\noalign{\smallskip}
BD+37 1541	&	A	&	f0:\,V	&		&		&		&	New F+M binary	\\	
Karmn J06353865+3751139 B	&	B	&	m2.5\,V	&	3.88	&	201.5	&	848.7	&		\\	\noalign{\smallskip}
LSPM J0651+1845	&	A	&	m4.5\,V	&		&		&		&	Confirmed known binary (FMR 83)	\\	
LSPM J0651+1843	&	B	&	m4.5\,V	&	111.72	&	150.6	&	7133.4	&		\\	\noalign{\smallskip}
1RXS J073138.4+455718	&	AaAb	&	M3\,V+m4.5\,V	&		&		&		&	New quadruple \\	
1RXS J073101.9+460030	&	B	&	M4.0\,V	&	431.39	&	296.0	&	24127.1	& B might be double, probably young ($\beta$ Pictoris) 	\\	
{[}SLS2012{]} PYC J07311+4556	&	C	&	m4\,V	&	307.80	&	266.3	&	17214.6	& 	\\	\noalign{\smallskip}
HD 61606 A	&	A	&	K3\,V	&		&		&		&	Confirmed known triple	\\	
HD 61606 B	&	B	&	K7\,V	&	57.90	&	112.7	&	815.2	&	No WDS entry but reported by \cite{Pov09}	\\	
BD-02 2198	&	C	&	M1.0\,V	&	3894.18	&	296.7	&	54822.6	&	C might be double	\\	\noalign{\smallskip}
1RXS J074948.5-031712	&	A	&	M3.5\,V	&		&		&		&	New triple	\\	
2MASS J07495087-0317194	&	B	&	m3.5\,V	&	1.93	&	266.3	&	32.8	&		\\	
2MASS J07494215-0320338	&	C	&	M3.5\,V	&	234.86	&	214.0	&	4002.1	& 	C might be double	\\	\noalign{\smallskip}
LP 209-28	&	``A''	&	m3:\,V	&		&		&		&	Disproved binary (KO 6)	\\	
LP 209-27	&	``B''	&	m4\,V	&	666.68	&	208.5	&	69836.7	&		\\	\noalign{\smallskip}
HD 77825	&	A	&	K2\,V	&		&		&		&	New K+M binary	\\	
1RXS J090406.8-155512	&	B	&	M2.5\,V	&	220.02	&	262.9	&	6026.0	&		\\	\noalign{\smallskip}
2MASS J13181352+7322073	&	A	&	m3\,V	&		&		&		&	New M+M binary	\\	
Gaia DR2 1688578285187648128	&	B	&	M3.5\,V	&	7.39	&	335.7	&	186.8	&		\\	\noalign{\smallskip}
HD 130666	&	A	&	G5\,V	&		&		&		&	New G+M binary	\\	
2MASS J14474531+4934020	&	B	&	m4.5\,V	&	29.54	&	336.6	&	3072.3	&		\\	\noalign{\smallskip}
TYC 2565-684-1	&	A	&	g1\,V	&		&		&		&	New G+M binary	\\	
2MASS J15080798+3310222	&	B	&	m3\,V	&	43.64	&	306.2	&	8644.3	&		\\	\noalign{\smallskip}
HD 134494	&	A	&	K0\,IV	&		&		&		&	Reclassified as sub-giant	\\	
BD+33 2548B	&	B	&	f9\,V	&	23.38	&	285.0	&	6455.7	& New m3-type companion to pair of evolved  \\	
Gaia DR2 1288848427727490048	&	C	&	m3\,V	&	5.85	&	180.3	&	1615.4	& solar-mass stars	\\	\noalign{\smallskip}
HD 149162	&	AaAbAc	&	K0\,Ve+k6\,V+m5\,V	&		&		&		&	Confirmed quintuple (LEP 79)	\\	
G 17-23	&	B	&	M3.0\,V	&	252.03	&	138.4	&	11405.7	&		\\	
LSPM J1633+0311S	&	C	&	D:	&	258.42	&	138.4	&	11694.8	&		\\	\noalign{\smallskip}
G 125-15	&	Aab	&	M4.5\,V+M5\,V	&		&		&		&	Confirmed triple (GIC 158)	\\	
G 125-14	&	B	&	M4.5\,V	&	45.78	&	347.4	&	1835.1	&		\\	\noalign{\smallskip}
LP 395-8 A	&	Aab	&	M3.0\,V+m0\,V	&		&		&		&		\\	
LP 395-8 B	&	B	&	m3.5\,V	&	1.92	&	355.5	&	56.6	&	New m9-type companion to trio of M dwarfs	\\	
Gaia DR2 1829571684884360832	&	C	&	m9:\,V	&	11.02	&	307.4	&	325.1	&		\\	\noalign{\smallskip}
HD 212168	&	A	&	G0\,V	&		&		&		&	Confirmed known quadruple (DUN 38, KO 5)	\\	
CPD-75 1748B	&	BaBb	&	k3\,V+	&	20.90	&	79.3	&	489.2	&		\\	
DENIS J222644.3-750342	&	C	&	M8\,V	&	264.82	&	128.9	&	6198.7	&		\\	\noalign{\smallskip}
SLW J2305+0613 A	&	``A''	&	M1.7\,V	&		&		&		&	Disproved triple (SLW 1300)	\\	
SLW J2305+0613 B	&	``B''	&	M3.2\,V	&	242.37	&	260.9	&		&	SLW 1300 is a visual pair	\\	
SLW J2305+0613 C	&	``C''	&	M3.7\,V	&	86.00	&	283.1	&	18656.7	&		\\	\noalign{\smallskip}
HD 221356	&	A	&	F7\,V	&		&		&		&	Confirmed known quadruple (KO 3, GZA 1)	\\	
2MASSW J2331016-040618	&	BC	&	M8.0\,V+L3.0\,V	&	451.70	&	261.7	&	11668.3	&		\\	
2MASS J23313095-0405234	&	D	&	L1\,V	&	12.46	&	221.6	&	321.9	&		\\	\noalign{\smallskip}
StKM 2-1787	&	A	&	K4\,V	&		&		&		&	Confirmed known binary (VYS 11)	\\	% No strictly physical in Montes+18, we confirm binarity
TYC 1174-955-2	&	B	&	M2.5\,V	&	5.78	&	165.1	&	215.5	&		\\\noalign{\smallskip}
\hline
\noalign{\smallskip}
\multicolumn{7}{l}{\footnotesize \textsc{Note}---Table 1 is published in its entirety in the electronic edition of {\it Research Notes}. A portion with fewer columns is shown here for}\\
\multicolumn{7}{l}{\footnotesize guidance regarding its form and content.}
\end{tabular}
\end{adjustwidth}
\end{table}
\newpage

\bibliography{cif02}{}
\bibliographystyle{aasjournal}

\end{document}